\documentclass[aps,twocolumn,superscriptaddress,prl,floatfix]{revtex4}
\usepackage{amssymb}
\usepackage{amsmath}
\usepackage{graphicx}
\usepackage{color}

\def\icrn{I_CR_N}
\def\mm{\mathrm{\mu}\mathrm{m}}
\def\vdc{V}
\def\idc{I}
\def\ic{I_{C}}
\def\rn{R_N}
\def\bc{B_{C}}
\def\ec{E_{C}}

\begin{document}
\title{Unconventional Josephson Effect in Hybrid Superconductor-Topological Insulator Devices}
\author{J.\ R.\ Williams}
\author{A.\ J.\ Bestwick}
\author{P.\ Gallagher}
\affiliation{Department of Physics, Stanford University, Stanford, CA 94305, USA}
\author{Seung Sae\ Hong}
\affiliation{Department of Applied Physics, Stanford University, Stanford, CA 94305, USA}
\author{Y.\ Cui}
\affiliation{Department of Material Science, Stanford University, Stanford, CA 94305, USA}
\affiliation{Stanford Institute for Materials and Energy Sciences, SLAC National Accelerator Laboratory, Menlo Park, California 94025, USA.}
\author{Andrew\ S.\ Bleich}
\affiliation{Geballe Laboratory for Advanced Materials, Stanford University, Stanford, CA 94305, USA}
\author{J.\ G.\ Analytis}
\author{I.\ R.\ Fisher}
\affiliation{Department of Applied Physics, Stanford University, Stanford, CA 94305, USA}
\affiliation{Stanford Institute for Materials and Energy Sciences, SLAC National Accelerator Laboratory, Menlo Park, California 94025, USA.}
\author{D.\ Goldhaber-Gordon}
\affiliation{Department of Physics, Stanford University, Stanford, CA 94305, USA}

\date{\today}

\begin{abstract}
We report on transport properties of Josephson junctions in hybrid superconducting topological insulator devices, which show two striking departures from the common Josephson junction behavior: a characteristic energy that scales inversely with the width of the junction, and a low characteristic magnetic field for suppressing supercurrent. To explain these effects, we propose a phenomenological model which expands on the existing theory for topological insulator Josephson junctions. 
\end{abstract}
\maketitle

The Majorana fermion, a charge-neutral particle that is its own antiparticle, was proposed theoretically almost 75 years ago~\cite{Wilczek09}. Electronic excitations in certain condensed matter systems have recently been predicted to act as Majorana fermions~\cite{Wilczek09}. One such system is a three-dimensional topological insulator (TI) where superconducting correlations between particles are introduced, producing a ``topological superconductor''~\cite{Fu08}.   When two superconductors are connected by a TI, the TI ``weak link'' superconducts due to its proximity to the superconducting leads. This produces a Josephson junction (JJ) but with several important distinctions compared to a conventional JJ, where the weak link is typically an ordinary metal or insulator.  Fu and Kane have predicted~\cite{Fu08} a one-dimensional (1D) mode of Majorana fermions at the interface between a conventional superconductor and a superconducting topological surface state.  Hence, JJs formed with a TI weak link are expected to have two 1D modes at the two superconductor-TI interfaces [arrows in Fig. 1(a)], which fuse to form a 1D wire of Majorana fermions [shown in purple in Fig. 1(a)] running along the width of the device~\cite{Fu08}. The energy spectrum of these Majorana fermions is characterized by states within the superconducting gap, which cross at zero energy when the phase difference $\varphi$ between the two superconducting leads is $\pi$.

\begin{figure*}[t!]
\center \label{fig1}
\includegraphics[width=5 in]{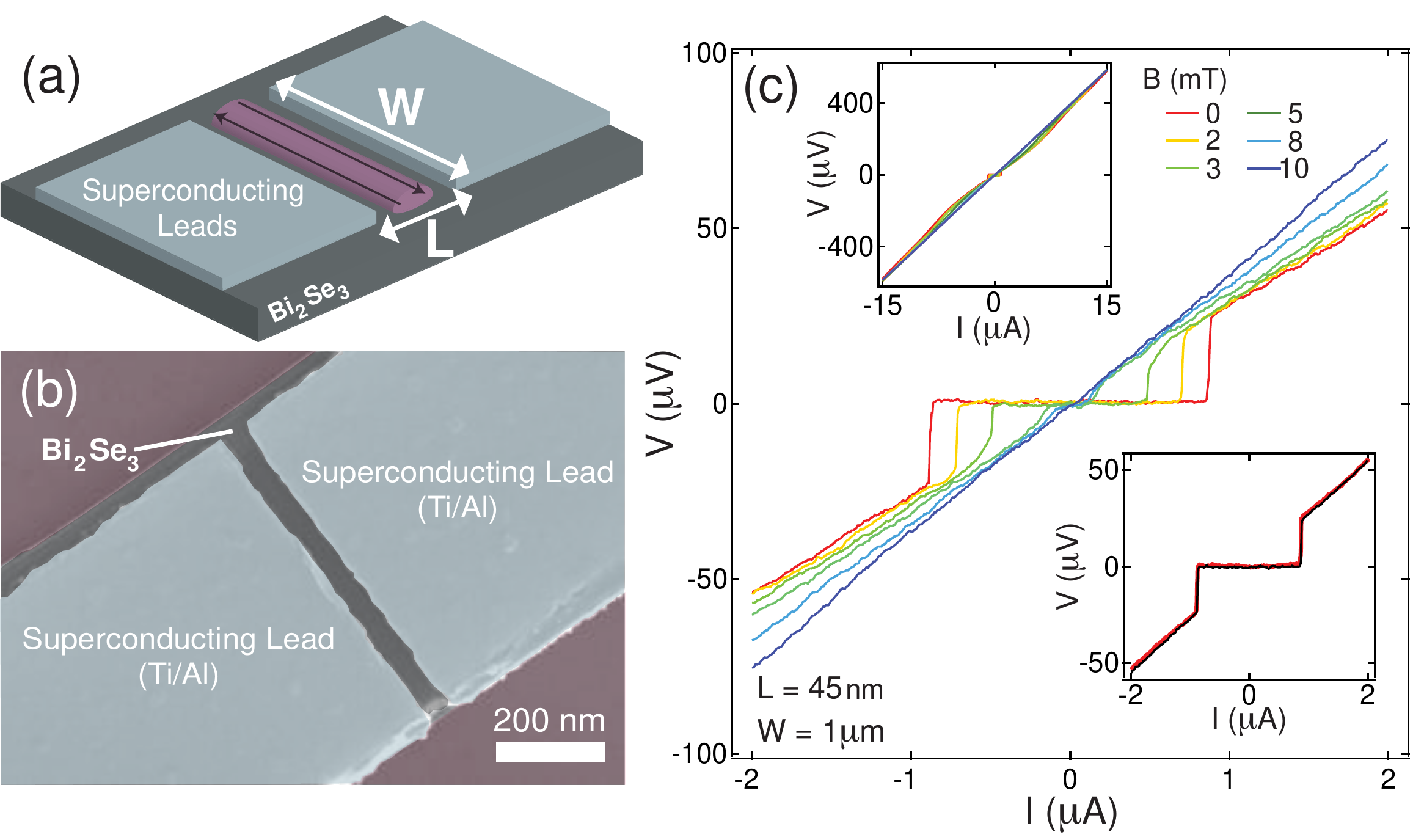}
\caption{(a) Schematic of a topological insulator Josephson junction. Two superconducting leads are patterned on top of Bi$_2$Se$_3$ forming a junction with length $L$ and width $W$. Along the width of the device, a one-dimensional wire of Majorana fermions results (purple).  (b) Scanning-electron micrograph of a device similar to the ones measured in this Report. (c) (main) $\vdc$ vs. $\idc$ for a devices of dimensions ($L,W$)=(45\,nm, 1\,$\mm$) for $B$=0, 2, 3, 5, 8, 10\,mT and at a temperature of 12\,mK. At $B$=0, $\ic$ is 850\,nA, which is reduced upon increasing $B$. For this device, the product $\icrn$=30.6\,$\mu$V, much lower than theoretically expected for conventional JJs. (upper-left inset) $I-V$ curves overlap for all values of $B$ at $\vdc \ge 2\Delta$/e$\sim$300\,$\mu$V. (lower-right inset) Sweeps up (red) and down (black) in $\idc$ show little hysteresis, indicating that the junction is in the overdamped regime.}
\end{figure*}

To probe this exotic state, recent experiments have investigated transport in TI JJs, finding good agreement with conventional JJ behavior~\cite{Sacepe11, Zhang11, Wang12, Qu11, Veldhorst11}. Two characteristic properties are typically reported for JJs. The first is the product $\icrn$, where $\ic$ is the critical current and $\rn$ is the normal state resistance. $\icrn$ should be of order $\Delta/e$ (where $\Delta$ is the superconducting gap of the leads and $e$ is the charge of the electron) and independent of device geometry~\cite{Tinkham96}. The second characteristic property is the ``Fraunhofer-like''  magnetic diffraction pattern, i.e. the decaying, oscillatory response of the supercurrent to the magnetic field $B$, applied perpendicular to the flow of the supercurrent. The first minimum in $\ic$ should occur at $B=\bc$, when one quantum of flux $\Phi_0=h/2e$ (where $h$ is Planck's constant) is passed through the area of the device. Recent reports on TI JJs~\cite{Qu11, Veldhorst11} match this expectation. 

In this Letter we report on transport properties of nanoscale Josephson junctions fabricated using Bi$_2$Se$_3$ as the weak link material. The main experimental results of this Report are two departures from conventional Josephson junction behavior in these devices: a small value of $\icrn$ that scales inversely with the width of the junction; and a value of $\bc$ that is $\sim$5 times smaller than that expected from the device area. Neither of these results is predicted or previously seen for conventional JJs nor TI JJs. To explain these experimental observations, we propose a two-fold phenomenological extension to the model in Ref.~\cite{Fu08}, with both extensions arising from accounting for confinement along the length of the 1D Majorana wire.

\begin{figure}[t!]
\center \label{fig2}
\includegraphics[width=3 in]{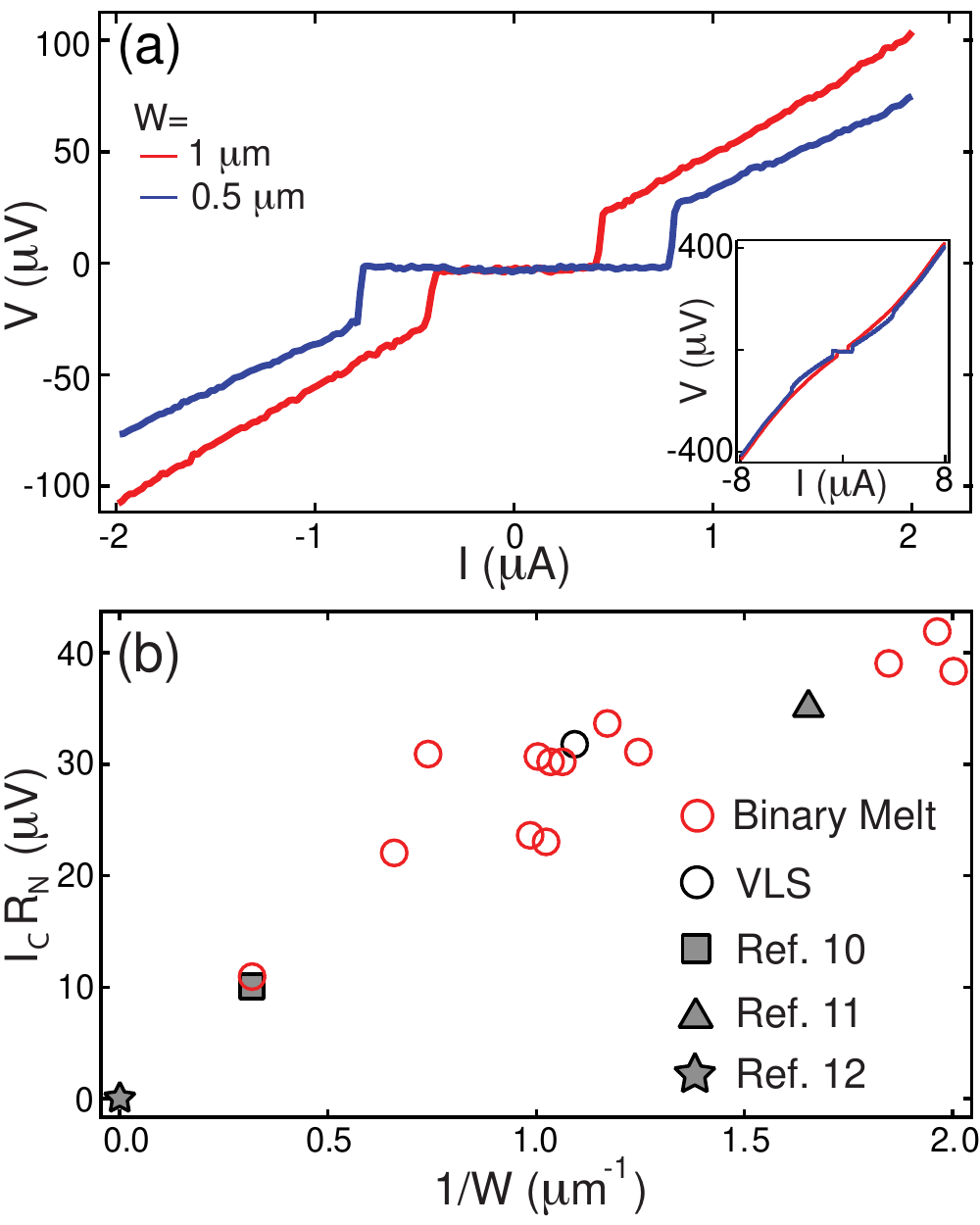}
\caption{(a), A comparison of two devices with similar $\rn$ (56.1 and 51.5\,$\Omega$) and different widths $W$, 0.5 and 1\,$\mm$. The device with $W$=0.5\,$\mm$ exhibits a larger $\ic$, in contrast to conventional JJs, where similar resistances lead to similar values of $\ic$. (b), $\icrn$ vs. $1/W$ for all 14 devices (synthesized via two methods: a binary melt and VLS) showing the general trend of $\icrn \propto 1/W$. In addition, $\icrn$ data points from Ref.~\cite{Sacepe11, Zhang11, Wang12} (grey symbols outlines in black) are shown in comparison to the results of this Letter.}
\end{figure}

To investigate the properties of JJs with TI weak links, junctions of lengths $L$ between 20 and 80\,nm and widths $W$ between 0.5 and 3.2\,$\mm$ were fabricated via electron-beam lithography and sequential deposition of Ti followed by Al to form electrical leads~\cite{SuppInfo} [Fig. 1(b)].  The DC response for a ($L$,$W$)=(45\,nm, 1\,$\mm$) junction at a temperature of 12\,mK is shown in Fig. 1(c), where the DC voltage ($\vdc$) is plotted as a function of the applied DC current ($\idc$).  At $B=0$, a typical DC Josephson response is observed (red curve): for $|\idc|$ $\leq$ $\ic$=850\,nA, $\vdc$=0 and a supercurrent flows. Applying $B$ perpendicular to the top surface of the Bi$_2$Se$_3$ reduces $\ic$ until $B$=10\,mT when the superconducting leads are driven normal and the I-V curve becomes linear. For $\idc > \ic$, there is an excess current due to Cooper pairs leaking into a low-barrier junction~\cite{Flensberg88}; this excess decreases with $B$. For $\vdc \ge$ 2$\Delta$/e$\sim$300\,$\mu$V, all curves fall on top of each other [upper left inset of Fig. 1(c)] for all values of $B$. Absence of hysteresis [lower right inset of Fig. 1(c), indicates that the junction is overdamped, consistent with calculations~\cite{SuppInfo}.  $\rn$ for this device is 35\,$\Omega$ and $\icrn$=30.6\,$\mu$V. Measurements of $\rn$ were carried out above the superconducting transition temperature of the leads in a four-terminal geometry, eliminating the resistance of the cryostat lines, but not the contact resistance between Ti/Al and Bi$_2$Se$_3$, which varies from device to device without apparent correlation to geometry or effect on $\icrn$ product. Theory~\cite{Likharev79} for diffusive or ballistic weak links predicts $\icrn$ to be 281\,$\mu$V or 427\,$\mu$V, respectively, an order of magnitude higher than our measurements. As a control experiment, a device fabricated similarly to the TI JJs, except with a 75\,nm-thick graphite weak link in place of Bi$_2$Se$_3$, has $\icrn$=244\,$\mu$V~\cite{SuppInfo}, much closer to theoretical predictions. This suggests that something in the sample rather than the measurement setup reduces the values of $\icrn$. 

\begin{figure}[t!]
\center \label{fig3}
\includegraphics[width=3 in]{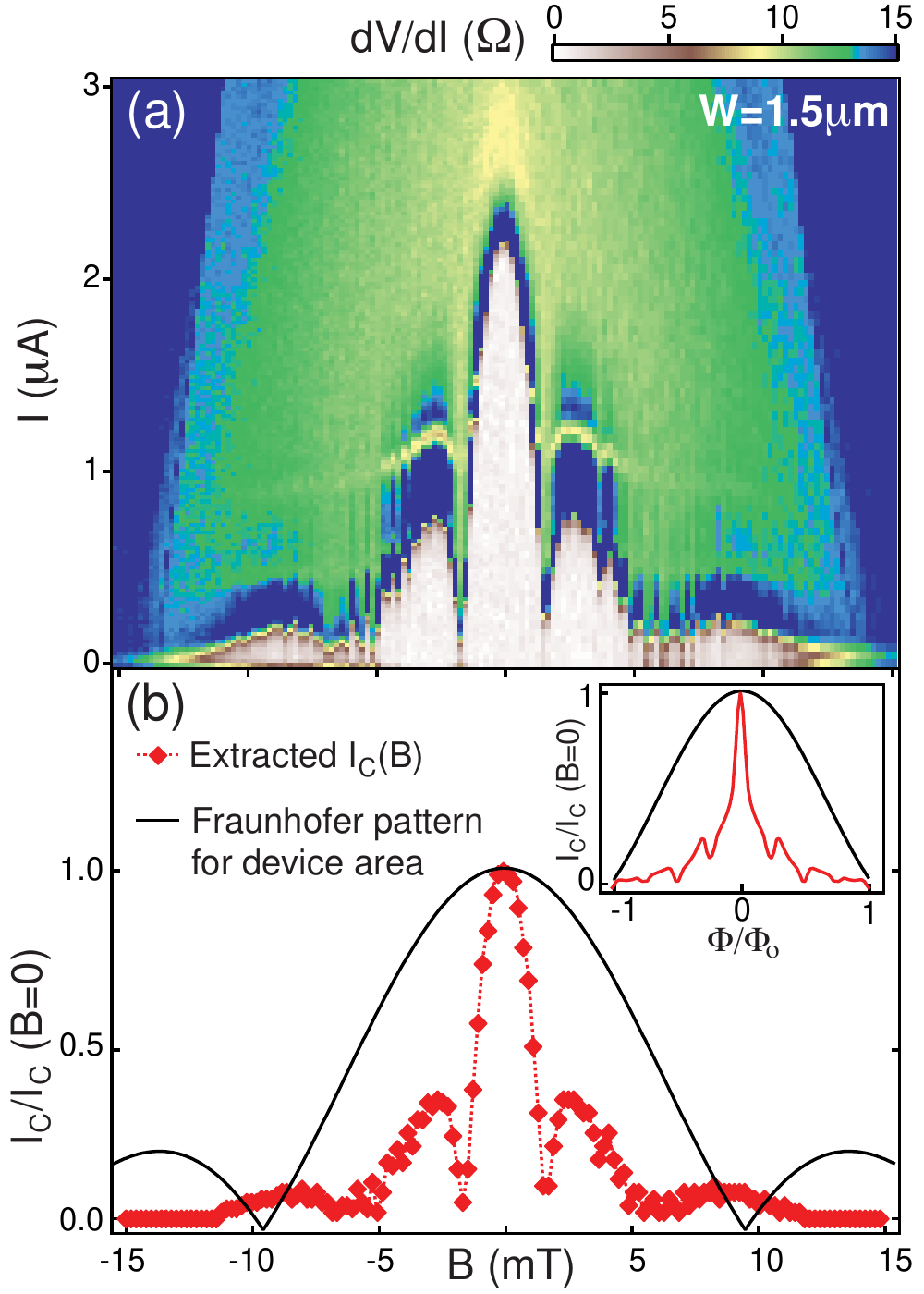}
\caption{(a), Differential resistance $dV/dI$ as a function of $B$ and $\idc$ showing an anomalous magnetic diffraction pattern for a $W$=1.5\,$\mm$ junction. Two features are of note: a smaller than expected value of $\bc$ at 1.70\,mT and a nonuniform spacing between minima at values $B$=1.70, 6.50, 11.80\,mT. (b), (main) $\ic\,(B)$ (red) extracted from $dV/dI$ in (a) is compared to the expected Fraunhofer pattern for the junction (black) where a reduction of the scale of the pattern and the nonuniform spacing are evident. (inset) A comparison of the simulated Frauhofer pattern for a sinusoidal (black) and an empirically-determined, peaked (red) CPR. The narrowing of the diffraction pattern and the aperiodic minima observed in (a) are captured this CPR.}
\end{figure}

Further insight into the nature of transport in TI JJs is found by investigating the width dependence of the characteristic quantity $\icrn$. A comparison of two junctions with $\rn$\,=\,56.1 and 51.5\,$\Omega$, and $W$=1\, and 0.5\,$\mm$, (both $L$=50\,nm) is shown in Fig. 2(a), where the 0.5$\mm$ device has roughly twice the critical current. The mismatch of the \emph{I-V} curves above $\ic$ is due to the excess current mentioned in connection with Fig. 1(c) (inset), which is typically of order $\ic$~\cite{Likharev79} and hence larger in the $W$=0.5\,$\mu$m device. The two curves approach each other as $\vdc$ approaches 2$\Delta/e$ [inset Fig. 1(c)].  The values of $\icrn$ for all 14 devices we measured that superconduct are shown as a function of 1/$W$ in Fig. 2(a). The trend is clear: a larger $W$ produces a smaller $\icrn$. With benefit of hindsight, results of some previously reported experiments on TI JJs are consistent with $\icrn$ being related to 1/$W$, and these are plotted alongside our data in shaded grey shapes outlined in black [Fig. 2(b)].  Specifically: in narrow topological insulator nanowires $\icrn$ is relatively high (triangle)~\cite{Zhang11}, though still well below predictions; for intermediate values of $W$ similar to ours, $\icrn$ is low (square)~\cite{Sacepe11}; in very wide junctions no supercurrent is observed at all (star)~\cite{Wang12}. To account for the different superconducting material used for the contacts, the value of $\icrn$ was scaled by the ratio of the superconducting gap of Aluminum to superconducting gap of the material used in Ref.~\cite{Wang12} (Indium) and Ref.~\cite{Zhang11} (Tungsten). Naively, $\icrn \propto \Delta$; also in the model we will introduce later, $\icrn \propto \Delta$ though with a smaller geometry-dependent prefactor.

The last characteristic response of TI JJs considered in this Letter is the magnetic diffraction pattern (MDP); our devices display an atypical relationship between $\ic$ and $B$. Fig. 3(a) shows the differential response $dV/dI (B,\idc)$ for a $(L,W)$=(55 nm, 1.5\,$\mu$m) device. Two phenomena are of note: $B_{C}$ is 5 times smaller than expected from the known device area and the shape of $\ic (B)$ deviates from a typical Fraunhofer pattern. The area of the devices is calculated as $W\ast(L+2\lambda_L$), where $\lambda_L$=50\,nm is the dirty London penetration depth for aluminum~\cite{SuppInfo}. The extracted $\ic(B)$ is shown in Fig. 3(b) (red) and compared to the simulated Fraunhofer pattern (black) for the device area~\cite{SuppInfo}. $B_{C}$ for this device is 1.70\,mT, whereas it should be 9.3\,mT, based on the device area measured from a scanning electron micrograph. We have measured a smaller-than-expected value of $\bc$ in all our devices. The three minima in $\ic$ on each side of $B=0$ are unequally spaced, occurring at $B$=1.70, 6.25, and 11.80\,mT. Even if the effective area of the junction were larger for unknown reasons, fitting the central peak to a Fraunhofer pattern would produce minima at 1.7, 3.4, and 5.1\,mT, different from what is observed. The graphite control device exhibits a more conventional MDP~\cite{SuppInfo}, with the first minimum close to the expected field.  

We have been unable to explain these experimental observations using known phenomena of conventional JJs, such as Pearl effects, flux focusing, and many others. It is not uncommon to observe reduced values of $\icrn$ in conventional JJs because of poor electric contact to the superconductor, thermal fluctuations or activation, or an extra normal channel that does not participate in supercurrent~\cite{Tinkham96}. Nor is it uncommon to have the first minimum of the MDP not at the expected field, because of flux focusing or nonuniform current distribution~\cite{Barone82}. Even considering all these effects, and others, as discussed in detail in the Supp Info, we are not able to account for such large deviations from naive expectations, with consistent behavior over many devices.  We therefore instead attempt to account for the effects seen in our Bi$_2$Se$_3$ devices in the framework of the model in Ref.~\cite{Fu08}. Since the original proposal did not consider our exact geometry or measurement, we propose a two-fold phenomenological extension to the model in Ref.~\cite{Fu08}: we do not claim to have proven that this phenomenological picture is correct, but since it accounts in an economical way for some of our striking observations we offer it as a spur to further theoretical and experimental work on this system. 

First we take into account confinement along the 1D Majorana wire, quantizing its energy levels at multiples of $\ec=h\nu_{ex}/2l$, with $\nu_{ex}$ the velocity of the carriers in the wire and $l$ the length of the wire.  In the present devices, the length of the wire is either the width $W$ of the JJ or, if the Majorana modes exist all the way around the TI flake, 2$W$+2$t$ (where $t \ll W$ is the thickness of the flake), hence $\ec \propto h\nu_{ex}/2W$.  The effect of this quantization on the energy levels is shown in Fig. 4(a). If the $E=0$ state [Fig. 4(a), solid purple dot] is topological in nature, i.e. it is a neutral Majorana mode, such confinement should not affect its existence nor change its energy from zero~\cite{Kitaev01}. The continuum of energy levels at $E \neq 0$, not protected from perturbations, is quantized in multiples of $E_C$ [Fig. 4(a), empty purple dots where only the first non-zero energy modes are shown for clarity]. 

The second extension of Ref.~\cite{Fu08} is to postulate the supercurrent is dictated by the physics of the junction near the zero-energy ($E=0$) crossings, whereas when higher-energy modes can be accessed (i.e. when $I R_N \geqslant E_c/e$) the transfer of carriers from one lead to the other is dissipative. Thus $\icrn$ set by the energy scale of confined modes along the width of the junction, rather than by $\Delta/e$: $\icrn \propto \ec/e = h\nu_{ex}/2eW$. There is disagreement in the literature on the relationship between $\nu_{ex}$ and $\nu_{F}$. However, in both available predictions the relationship is of the form $\nu_{ex}=\nu_{F}(\Delta/\mu)^n$, where $\mu$ is the chemical potential and n=2 in Ref.~\cite{Fu08} and n=1 Ref.~\cite{Titov07}. As shown in~\cite{SuppInfo}, the result of~\cite{Fu08} would give energies far too low to account for our observations, so we arrive at the relationship $\icrn \propto \Delta$ as in conventional JJs, but with different constants of proportionality, some relating to the geometry of the device.

The confinement would also has an effect on the current-phase relation (CPR), which determines the supercurrent through the device as a function of $\varphi$. The supercurrent enabled by the $E=0$ state occurs at $\varphi=\pi$,  producing a sharp peak in the CPR at $\varphi=\pi$ [Fig. 4(b)], in contrast with the established sinusoidal CPR for conventional JJs~\cite{Tinkham96}, or a doubled-period sinusoid predicted for TI JJs~\cite{Fu09}. The locations and shapes of the peaks in the CPR depend on the details of the energy spectrum of the bound electron-hole pairs,  as discussed further in~\cite{SuppInfo}.

\begin{figure}[t!]
\center \label{fig4}
\includegraphics[width=3 in]{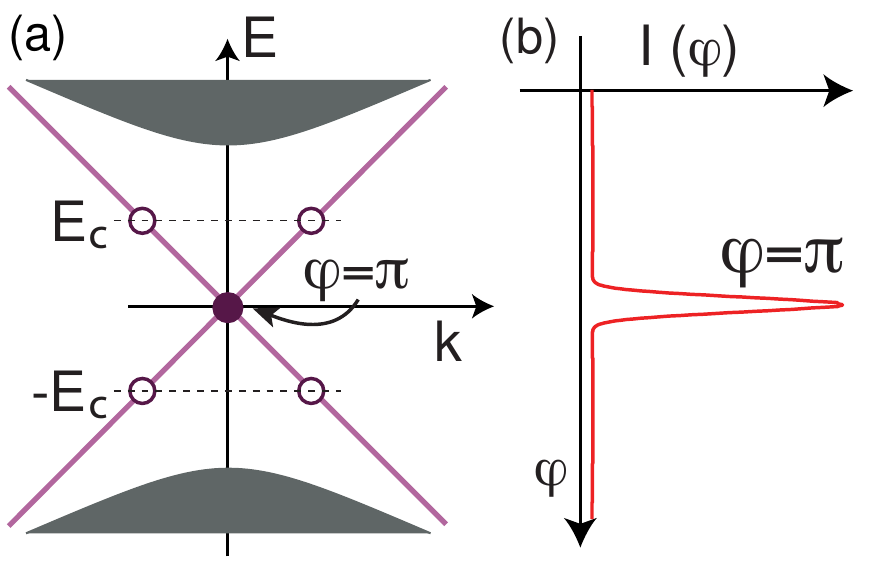}
\caption{(a) Energy levels near $\varphi=\pi$ before momentum quantization along $W$ (purple line) and after, where the topological state remains at $E=0$ (solid purple circle) and the first quantized energy level at the value $\ec$ (empty purple circles). (b) Current-phase relation resulting from momentum quantization, producing an anomalous peak at $\varphi=\pi$. The location and shape of the peaks current-phase depend on the details of energy spectrum of the Andreev bound pairs and in ~\cite{SuppInfo} we consider several possible scenarios for this spectrum. }
\end{figure}

As described above, the result of confinement is to separate in energy the $E=0$ 1D modes (neutral Majorana modes) from the $E \ne 0$ (charged) 1D modes. A supercurrent can pass through a charged 1D mode, but the critical current is strongly suppressed by interactions between charges~\cite{Fazio95, Maslov96}. Thus the supercurrent associated with the zero crossing should be larger than that associated with the charged
modes at higher energy.. In our experiment the lowest-energy charged modes are accessed when the current associated with the zero mode is $\sim$1\,$\mu$A. We estimate that the charged modes cannot carry this much supercurrent. Hence the charged modes in our devices act as resistors carrying current but not supercurrent. When the charged modes become energetically accessible (i.e. for energies $\ge \ec$) they shunt the junction, with an expected 1D-charged-mode shunt resistance greater than $h/4e^2$~\cite{Kane92}. Renormalization group calculations show that for this value of the shunt resistance supercurrent shuts off, and the JJ behaves as a metal~\cite{Schon90}. In this model, any additional supercurrent through the bulk (which might be expected in existing TIs, given the finite bulk conductivity) also ceases when the shunt resistance of the surface become energetically accessible. 

Additional peaks in the CPR at certain values of $\varphi$, suggested by our phenomenological model as noted above, produce a narrowing in the MDP as observed in Fig. 3. We note that an anomalous, peaked CPR has been theoretically predicted for a different device geometry, also a result of the presence of Majorana fermions~\cite{Ioselevich11}.  In each case, one sharp feature in the CPR occurs for each zero-energy crossing of states in the gap. For a single peak in the CPR, only a single, $B$=0 maximum in the MDP is possible for $|\Phi| < \Phi_0$~\cite{SuppInfo}. The existence of multiple oscillations in $\ic(B)$ for $|\Phi| < \Phi_0$ strongly suggests the presence of multiple peaks in the CPR -- possibly a result of coupling to fermionic modes in the device that create additional zero-energy crossings~\cite{Law11}.  Through simulation, we are able to show that a smoothly-varying CPR cannot capture our results~\cite{SuppInfo} and only a CPR with peaks can create a MDP even coarsely resembling those observed in our devices, i.e. we cannot describe our MDPs using conventional effects like flux focusing or nonuniform supercurrent distribution alone. 

When considering the MDP, it is important to note that known topological insulators including Bi$_2$Se$_3$ and Bi$_2$Te$_3$ have contributions to conductance from both the bulk and the anomalous surface state; both must taken into account when considering supercurrent flow and the CPR of the Josephson junction. A MDP derived from a peaked CPR (from the surface state) added to a conventional, sinusoidal CPR from the bulk with 1/5 of the amplitude of the surface state~\cite{SuppInfo} is compared to the  typical Fraunhofer pattern (black) in the inset of Fig. 3(b). Some, but not all, of the features observed in the experiment are captured by this CPR. Importantly, two features are captured by this peaked CPR: the MDP is narrowed, and non-uniformly distributed minima occur at $\Phi_0/4, \Phi_0/2,$ and $\Phi_0$~\cite{SuppInfo}, near the aperiodic structure of the minima seen in the experiment.  In our materials, a significant effort has been made to reduce the bulk contribution to conductance~\cite{Analytis10,Peng10}.  A systematic investigation of the effect of a bulk supercurrent contribution to the MDP is performed in Ref.~\cite{SuppInfo}, where it is found that for roughly equal contributions of the surface and bulk to the CPR, a more conventional MDP results with minor deviations from a Fraunhofer pattern. For example, deviations from a Fraunhofer pattern generated in the simulations of Ref.~\cite{SuppInfo}, such as a triangular-shaped central node, can be observed in Ref.~\cite{Qu11}.

\textbf{Acknowledgments:} We thank Shaffique Adam, Malcolm Beasley, John Clarke, Liang Fu, Sophie Gu\'eron, Jedediah Johnson, Patrick Lee, Chris Lobb, Joel Moore, Chetan Nayak, Xiaoliang Qi, and Victor Yakovenko for valuable discussions. The work was supported in part by the Keck Foundation. J. R. W. acknowledges support from the Karl van Bibber Postdoctoral Fellowship. A. J. B. from an NDSEG Fellowship. J. G. A., A. S. B. and I. R. F. are supported by the DOE, Office of Basic Energy Sciences, under contact DE-AC02-76SF00515

\newpage

\section{Supplementary Information}

\subsection{Materials and methods}
Single crystals, grown by the binary melt method, of (Bi$_{1-x}$Sb$_x$)$_2$Se$_3$ were synthesized by slow cooling a binary melt of Bi (99.9999$\%$), Sb (99.999$\%$) and Se (99.9999$\%$) starting materials, mixed in the ratio 0.33:0.05:0.62. The actual amount x of Sb in the single crystals is approximately 0.01, as measured in microprobe analysis. Bulk single crystals with dimensions 1$\times$1$\times$0.1mm$^3$ showed a non-metallic resistivity at low temperatures with a bulk carrier density of 7$\times$10$^{16}$ cm$^{-3}$ as deduced by Hall and Shubnikov-de Haas analysis. 

Bi$_2$Se$_3$ nanowires, synthesized via the vapor-liquid-solid (VLS) method, were grown in a 12-inch horizontal tube furnace with a quartz tube. Bi$_2$Se$_3$ source powder (99.999$\%$, from Alfa Aesar) was placed in the center of the furnace (540\,$^{\circ}$C). The growth substrate, a Si wafer with a thermally-evaporated, 10\,nm Au film, was placed at a downstream, lower temperature zone (350\,$^{\circ}$C). High purity Ar gas delivers vapor from the source materials to the growth substrate at 130\,sccm for 2 hours with 1\,Torr pressure.

Samples of 50-100\,nm thickness were prepared by mechanical exfoliation of the above two starting material, using a method similar to exfoliation of graphene~[S1].  Prior to sample exfoliation, SiO$_2$(300\,nm)/Si wafers were cleaned in acetone and isopropanol.  HMDS was deposited on the SiO$_2$ surface in efforts to reduce the amount of H$_2$O contamination. Samples were exfoliated using Nitto-Denko tape and immediately after exfoliation PMMA 950 A4 (Microchem Crop.) was spun on the chip for 60\,secs at 4000 rpm, followed by a baked for 30\,mins at 180\,$^{\circ}$C on a hotplate. After electron beam exposure of the resist and development in MIBK/IPA 1:3, a 3\,min UV-ozone was used to remove residual PMMA in the exposure contact areas. Ion milling of the exposed contact areas was performed to obtain low contact resistance to the Bi$_2$Se$_3$ flakes. The recipe of the ion mill is a 300\,V beam voltage, 30\,mA beam current producing an ion current density of 0.1\,mA/cm$^2$, 60\,V acceleration, 12\,sccm argon flow, and a 10\,secs exposure. Immediately following ion milling, the samples were loaded into an e-beam evaporator equipped with oil-free pumps, and Ti/Al (3\,nm and either 60\,nm or 100\,nm) were evaporated at a rate of 1\,${\mathrm{\AA}}$/s onto the sample at a pressure of $\sim$5$\times$10$^{-8}$\,Torr at the start of each evaporation. 

Samples were measured in a $^3$He/$^4$He dilution refrigerator with a base temperature of 12\,mK. To reduce the effects of thermal radiation, three stages of filtering were employed: room-temperature LC $\pi$-filters, a 3-pole RC filter anchored to the mixing chamber designed to filter in the frequency range of 1\,MHz to 5\,GHz~[S2], and 10\,k$\Omega$ ``on-chip'' resistors used in conjunction with the inline capacitance of the  twisted-pair wiring to produce a low-pass, RC-filter of cutoff frequency $\sim$50\,KHz.  All experiments were performed in an electrically- and acoustically-shielded room. Device DC resistance and differential resistance were measured using a standard DC and lock-in technique. The applied AC current was kept low enough such that the AC voltage across the device never exceeded 500\,nV, below the base temperature of the fridge k$_B$T$\sim$1\,$\mu$eV. To obtain accurate magnetic field values, a low-noise Yokogawa 7651 was used to supply the current to the superconducting magnet, producing a $<$0.3\,$\mu$T noise level in the magnet. 

The transport properties of the exfoliated binary melt samples were characterized by making a device with contacts in a Hall-bar geometry and Bi$_2$Se$_3$ thickness of 100\,nm. The extracted density, mobility and diffusion constant $D$ of these samples are 7.8$\times$10$^{17}$\,cm$^{-3}$, 1950\,cm$^2$/Vs, 0.02\,m$^2$/sec, respectively.  We note that these values reflect merged contributions from both the surface and bulk, where it is expected that the surface electrons have a much higher mobility~[S3]. Hence, the values for mobility and $D$ are a lower bound for the values of the surface electrons. 

To extract the superconducting and normal state properties of the leads, four samples were prepared with the same recipe as the leads of the JJ devices: Ti/Al of thickness  3\,nm/(60, 100\,nm), both of  $(L,W)$=(10,1)\,$\mm$. Each thickness of Al was placed either directly on the SiO$_2$ substrate or on top of a Bi$_2$Se$_3$ flake. Characterization of these test samples was performed via a four-terminal resistance measurement at room temperature (290\,K), 4\,K and  base temperature of the dilution refrigerator (12\,mK). Little variation was found in the superconducting properties of the Al patterned on the SiO$_2$ substrate and on Bi$_2$Se$_3$. The resistivity of the samples at room temperature and 4K were 3.0$\times$10$^{-8}$\,$\Omega$-m and 0.3$\times$10$^{-8}$\,$\Omega$-m, respectively. The critical currents and magnetic fields for the 100\,nm(60\,nm)-thick Al-films were 488\,$\mu$A(208\,$\mu$A) and 11(13)\,mT respectively.

\subsection{Characteristic lengths and regimes for JJs}
There are many different regimes in JJs that produce different behaviors. The regime applicable to the device is determined by the length of the junction $L$ relative to five intrinsic length scales: coherence length of the Al leads $\xi_{Al}$=1.6\,$\mm$; the superconducting coherence length of the metallic (Bi$_2$Se$_3$) weak link $\xi_{N}$, and three properties of the (Bi$_2$Se$_3$) weak link unrelated to the superconductor: the thermal length $\ell_T$, the mean-free path $\ell_e$ and the phase-coherence length $\ell_{\varphi}$. Several of these lengths depend on the diffusion constant. We use the merged value for D as noted above (see below for more comments on this). $\xi_{N}$ can be calculated from the diffusion constant by $\sqrt{\hbar D/\Delta}$, where $\Delta$ is the superconducting gap of Al, which we calculate from the BCS equation $\Delta$=1.76\,$k_BT_C$, which for a $T_C$ of 1\,K (a typical value for our leads) is 151\,$\mu$eV, This produces a value for $\xi_{N}$ of $\sim$\,280\,nm. $\ell_T=\sqrt{\hbar D/2 \pi k_B T}$, which at 12\,mK is 1.3\,$\mm$. $\ell_e$ can be extract from D as D=1/2\,$\nu_F \ell_e$, giving $\ell_e$=80\,nm, where we have used the value of $\nu_F$=4.2$\times$10$^5$\,m/s from ARPES measurements~[S4]. $\l_{\varphi}$ was estimated from the half-width at half-max of the weak antilocalization correction to the longitudinal resistance (measured in the Hall bar) to be 650\,nm. We note that $\xi_N$, $\ell_T$, $\ell_e$ and perhaps $\ell_{\varphi}$ all represent lower bounds since we have used the average value of $D$ from the surface and the bulk. The $D$ for the surface electrons is expected to be higher.  

$W$ for the device should be compared to the Josephson penetration depth $\lambda_J=\sqrt{\Phi_{\mathrm{o}}/(2\pi\mu_{\mathrm{o}}j_C(2\lambda_L+L))}$, where $\lambda_L$ is the London penetration depth $\sim$ 50nm~[S5]. We place a lower bound on $\lambda_J$ by using the junction parameters that would make it the smallest, i.e. using the largest measured value of $j_C$ and the longest value of $L$: producing a value for $\lambda_J^{min}$ of 10\,$\mm$. This puts our junctions in the regime L$\sim$$\ell_e <  \xi_{N}, \xi_{Al}, \ell_{\varphi}, L_T$ and $W<\lambda_J$, i.e. the short junction, quasi-ballistic regime. In this regime, $\icrn$ is expected to be described by either the KO-1 or KO-2 theory~[S6] and should be independent of $W$: neither of these expectations matches our data.

\subsection{Supercurrent carried by the bulk}
The typical thickness $t$ for our devices is 75\,nm. The calculated $\xi_N$ ($\sim$\,280\,nm) is larger than this value, which suggests that both the surface and \emph{entire} bulk should experience the proximity effect. We expect, however, that the supercurrent carried by the bulk to be smaller than that of the surface for two reasons. First, the Cooper pairs have to travel a longer distance from left to right lead through the bulk than through the surface. We can estimate this additional length by assuming the Cooper pairs have to go on average a distance $t$/2 from the left lead to the bulk, then another $t$/2 going from the bulk to the right lead. This makes the effective length of the device for the bulk states to be $L_{bulk}=L+t$. The second is the lower mobility of the bulk, producing a smaller value of $D$. For the device in Fig. 3 of the main manuscript $L$=55\,nm and t=95\,nm, making $L_{bulk}$=150\,nm. To calculate the reduction of the critical current, we assume $D$ for the bulk states is 10 times smaller than the calculated value (a factor 12 has been experimentally determined~[S7] for the reduction of $D$ for the bulk), giving a $\xi_N^{bulk} \sim$ 65\,nm. This estimate yields a bulk critical current of order five times smaller than that of the surface state~[S6].

\subsection{Mechanisms for the reduction of $\icrn$}
We next consider noise as a mechanism for the reduction of $\icrn$. A typical reduction comes from thermal noise, where the effect on $\icrn$ depends on the type of junction (overdamped or underdamped). The type of junction is determined the quality factor $Q=\sqrt{2e\ic C/\hbar} \, \rn$, where $C$ is the capacitance of the JJ. We estimate $C$ to be 0.5pF using a parallel plate capacitor model between the entire area of the leads of the device and the degenerately-doped Si, yielding $Q$=1.2. This is neither overdamped or underdamped, but at an intermediate value. Numerical evaluation of the \emph{I-V} characteristics in this regime shows results similar to those of an overdamped junction~[S8]. The effect of thermal radiation on overdamped junctions has been calculated~[S8], and shown to cause a smearing of the transition between the superconducting and normal state. Since the transitions we measure are still very sharp, we can rule out thermal fluctuations being a source of the reduction of $\icrn$. We can also calculate what the effect of thermal fluctuations would be had the junctions been in the underdamped regime~[S5]: the effective thermal radiation temperature would have to be 3.4\,K to reduce $\icrn$ by the factors we observe. This high temperature is unphysical in our well-filtered setup. 

\begin{figure}[t!]
\center \label{figs1}
\includegraphics[width=3 in]{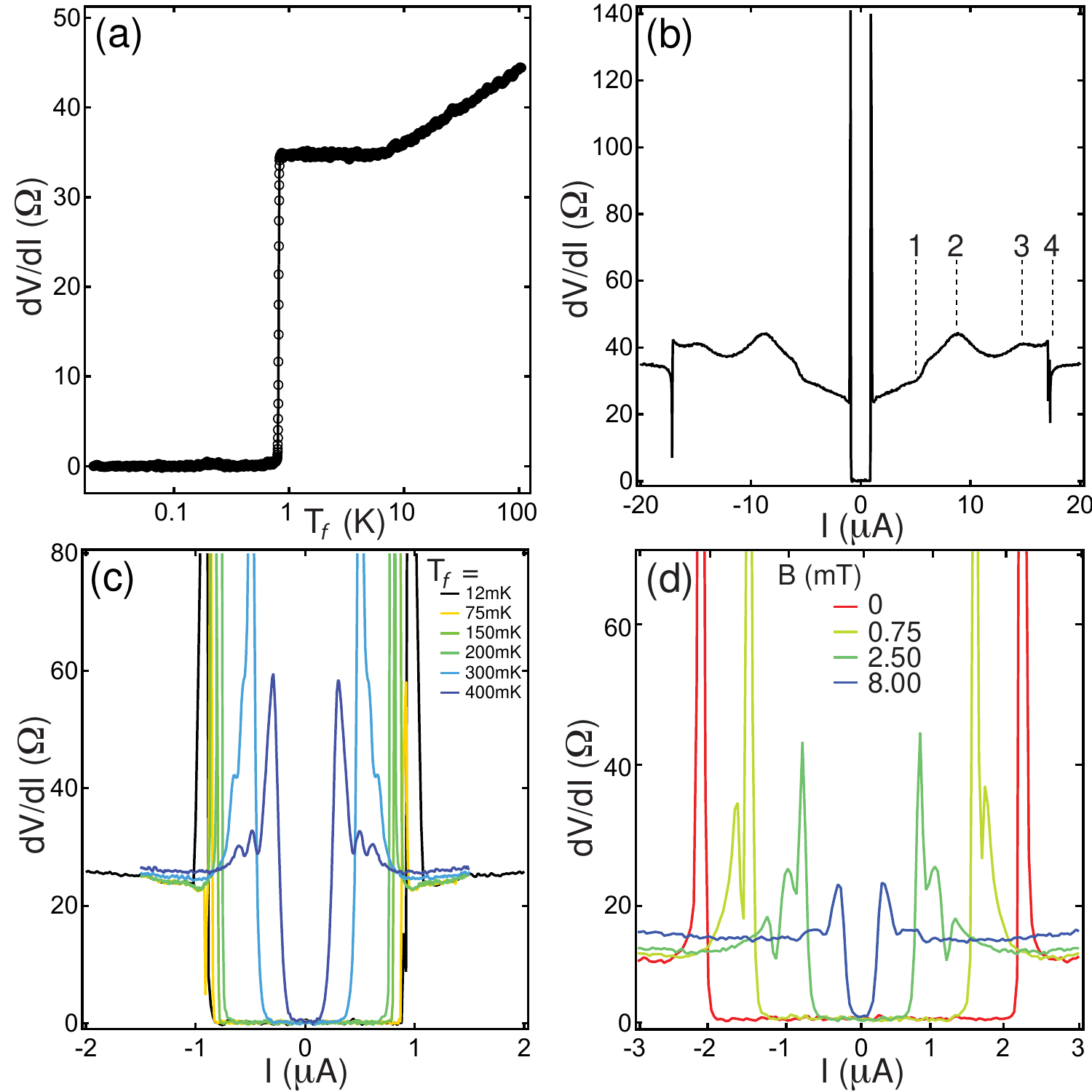}
\caption{(a) Differential resistance $dV/dI$ as a function of fridge temperature $T_f$ as $T_f$ is lowerd through $T_C$ of the leads. For this device, a sharp transition to the superconducting state is observed at $T_f$=850\,mK. (b) Additional features in $dV/dI$ as a function of $\idc$ are observed above $\ic$. (c) $dV/dI$ for values of $T_f$=12, 75, 150, 300, 400\,mK. Shoulder dips are observed in $dV/dI$ for the two temperatures 300 and 400\,mK. (d) $dV/dI$ extracted for B=0, 0.75, 2.50, 8.00\,mT from Fig. 3(a) of the main text. Shoulder dips are also apparent when the magnetic field is increased to 0.75\,mT}
\end{figure}

\subsection{Additional features in the Differential Response dV/dI as a function of T and B}
The differential response $dV/dI$ for a $(L,W)$=(50\,nm, 0.9\,$\mm$) device as a function of fridge temperature $T_f$ is shown in Fig. S1(a), where the device exhibits a sharp transition to $dV/dI$=0 below $T_f$=850\,mK of the leads. Above T$_C$, the devices shows a metallic temperature dependence of resistance. The differential response $dV/dI(\idc)$ of this junction, typical for all devices measured, at $T_f$=12\,mK is shown in Fig. S1(b).  Even aside from the transition from a superconducting state to a normal state, which produce large peaks at $\idc=\ic$, several features are evident and are indicated by labels 1-4. The first occurs in all devices as either a peak or a sharp rise in $dV/dI$ [like the one shown in Fig. S1(b)], at $\vdc$\,=\,200\,$\mu$V. The second, broad peak appears at $\vdc$\,=\,300\,$\mu$V, near 2$\Delta$/e of Al, where a peak in resistance is expected due to the quasiparticle contribution to the resistance from the leads~[S9].  The third and fourth peaks are both features occurring above 2$\Delta$ of the leads, occurring at $\vdc$\,=\,500\,$\mu$V and $\vdc$\,=\,655\,$\mu$V, respectively. Peaks above 2$\Delta$ have also been seen previously in TI JJs~[S10].

The effect of $T_f$ on $dV/dI$ is shown in Fig. S1(c). A typical trend is observed, where increasing $T_f$ reduces $\ic$. Additional ``shoulder'' dips appearing in $dV/dI$ are evident in the $T_f$=300\,mK and 400\,mK traces. On the 400\,mK trace, two dips are seen, at $\idc$=500\,nA and 620\,nA respectively. 

Additional shoulder features are also seen in $dV/dI$ above $\ic$ when a magnetic field $B$ is applied [Fig. S1(d)]. These are revealed in cuts of constant $B$ in $dV/dI(B,\idc$) for the device shown in Fig. 3 of the main text. At $B=0$, there are two peaks in $dV/dI$ at $\idc=\pm\ic$, and no other apparent features. At $B$=0.75\,mT, dips are evident at $\idc=\pm$1.6\,$\mu$A, $\vdc$=12.5\,$\mu$V above the resistive transition. The single dip occurs for 0.6$\le$$B$$\le$1\,mT. Two dips appear beginning at $B$=1\,mT and persist throughout the entire second and third lobes, although to a much weaker degree in the third lobe.  The double-dip is evident in the cut at $B$=2.5\,mT, occurring at $\idc \pm$=0.9 and 1.2\,$\mu$A and a value of $\vdc$=7.4 and 12.1\,$\mu$V. The dips at $B$=8\,mT have all but disappeared, remaining weakly at $\idc \pm$=0.5 and 0.9\,$\mu$A and $\vdc$=6.3 and 12.1\,$\mu$V. Currently, we have no explanation for these features, which appear in all devices measured.

\begin{figure}[t!]
\center \label{figs2}
\includegraphics[width=3 in]{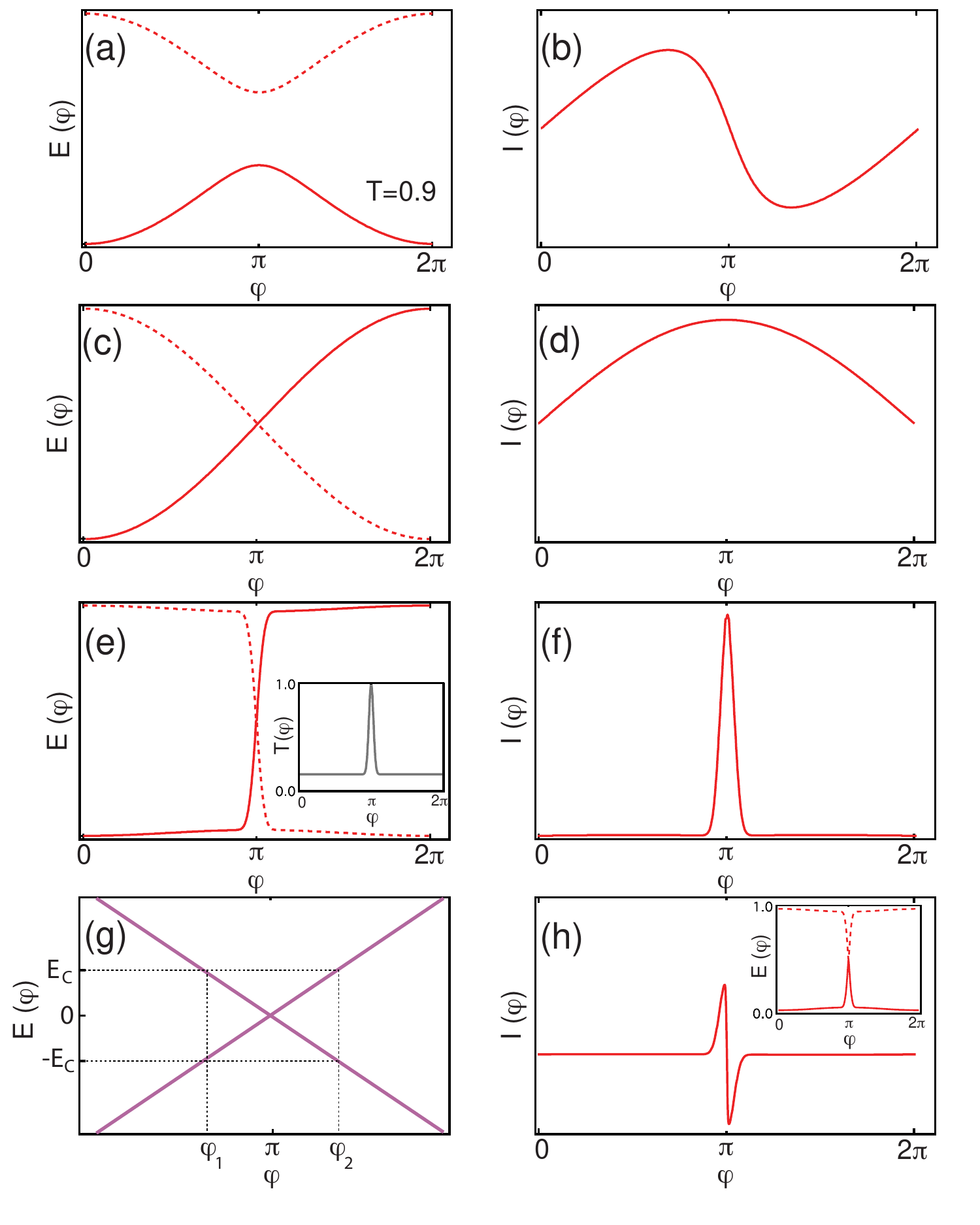}
\caption{The CPR can be determined from the Andreev bound state (ABS)  energy spectrum by taking the derivative of $E$ with respect to $\varphi$. Qualitatively shown is the ABS spectrum (a) and the corresponding CPR (b) for conventional, ballistic JJs with a transmission probability $T$=0.9. The same plots (c,d) are shown for $p+ip$ and TI (without the lateral confinement of our model) JJs. ABS spectrum and CPR (e,f) for the phenomenological model of the main text. The inset of (e) shows the dependence of $T$ on $\varphi$ in our model. The CPR of (f) is peaked at the values of $\varphi$ where the zero-energy crossings occur. (g) The width of $T$ in the inset of (e) and the corresponding CPR are set by $\ec$, the lateral confinement energy in the phenomenological model. (h) (inset) ABS spectrum similar to (e) except with a anti-crossing in the spectrum at $\varphi=\pi$. (main) CPR resulting from the ABS of the inset.}
\end{figure}

\subsection{Andreev Bound States and the Current-Phase Relation}
In this section we consider qualitatively the Andreev bound states (ABS) and the resulting current-phase relation (CPR). States in a JJ within the superconducting gap are associated with the coupled electron-hole pairs that are responsible for transferring Cooper pairs from one lead to the other~[S11]. These are known as ABS. To describe the CPR in our junction, we consider several different plausible energy spectra of ABS. The energy levels of conventional ABS are $E(\varphi) \propto \pm \Delta \sqrt{1-Tsin^2(\varphi/2)}$~[S11], where $T$ is the transmission probability of Cooper pairs from left to right lead. From the energy spectrum of the ABS, the current phase relation (CPR) can be calculated as $I(\varphi) \propto \partial E/\partial \varphi$~[S10]. The energy spectrum for the ABS in a conventional JJ with $T$=0.9 is shown in Fig. S2(a) for the occupied (solid) and unoccupied (dashed) ABS states. The corresponding CPR, calculated by differentiating the $E(\varphi)$ plot, is shown in Fig. S2(b), where a saw-tooth behavior is seen, occurring with a period of 2$\pi$. Note that for low values of $T$, this CPR becomes the conventional $sin(\varphi)$~[S11]. For $p+ip$ and TI JJs, the energy of the ABS becomes $E(\varphi) \propto \pm \Delta cos(\varphi/2)$, producing a zero-energy crossing at $\varphi=\pi$~[S12]. The $p+ip$ ABS spectra matches the conventional energy spectrum of ABS for $T=1$. The corresponding CPR --assuming a constant ABS occupancy rather than always being in the ground state, i.e. conserved fermionic parity -- is $I(\varphi) \propto sin(\varphi/2)$, resulting in an anomalous 4$\pi$-periodic CPR, predicted for $p+ip$ JJs~[S12] and TI~[S13] JJs.  The energy spectrum of the ABS, and the CPR, in this case are shown in Fig. S2(c,d).

In our phenomenological model, the transmission coefficient $T$ depends on $\varphi$, where the 1D charge modes at $\varphi \neq \pi$ have a low value for $T$ -- a result of the interactions in 1D~[S14] -- and the neutral modes have a much higher value for $T$. This $T(\varphi)$ produces a CPR that is peaked at values of $\varphi$ where the zero-energy modes occur.  Fig. S2(e) shows an ABS spectrum for a $T(\varphi)$ [inset of Fig. S2(e)], corresponding to $T(\varphi)$=1 for $\varphi=\pi$ and $T(\varphi)$=1/10 for $\varphi \neq \pi$.  The exact value for $T$ away from $\varphi = \pi$ can be changed without much change in the results . To connect the high- and low-$T$ regions, we assume that the width of the peak in $T(\varphi)$ near $\pi$ is set by the energy scale $\ec$, the energy associated with momentum quantization in the phenomenological model. This scenario is shown in Fig. S2(g), where the linearized $E(\varphi$) is calculated from Eq. 4 of Ref.~[S15].  Using a value of $\ec$=$e \icrn$=31$\,\mu$eV for the device in Fig. 3, values of 2.91\,rads (167$^{\circ}$) and 3.37\, rads (193$^{\circ}$) for $\varphi _1$ and $\varphi _2$ are obtained. $T$ then is modeled as a gaussian with a half max at the values $\varphi _1$ and $\varphi _2$. The calculated ABS spectrum for this transmission coefficient is shown in Fig. S2(e), where a transition through $E=0$ occurs on the scale of $\varphi _1$ and $\varphi _2$ around $\varphi=\pi$. The corresponding $I(\varphi)$ is shown in Fig. S2(f), where the $\varphi$-dependent $T$ produces a anomalous CPR that is peaked around the values of $\varphi$ where the zero-energy modes occur. This CPR is also 4$\pi$ periodic. A similar CPR was obtained in Ref.~[S16], where a peak in the CPR was also found to be correlated with the zero-energy crossing in the ABS. In Fig. S2(c) and S2(e), it was assumed that fermonic parity was conserved, resulting in a protected crossing of the two states (solid and dashed lines) at $\varphi=\pi$~[S17]. If the parity of the junction is not preserved, an anti-crosssing of these two states occurs [inset of Fig. S2(h)], producing a 2$\pi$-periodic CPR shown in Fig. S2(h). This 2$\pi$ periodicity will likely apply to any near-DC (as opposed to microwave) measurement. Nonetheless, below we initially try a 4$\pi$-periodic candidate CPRs before returning to 2$\pi$ periodicity.

\begin{figure}[t!]
\center \label{figs3}
\includegraphics[width=3 in]{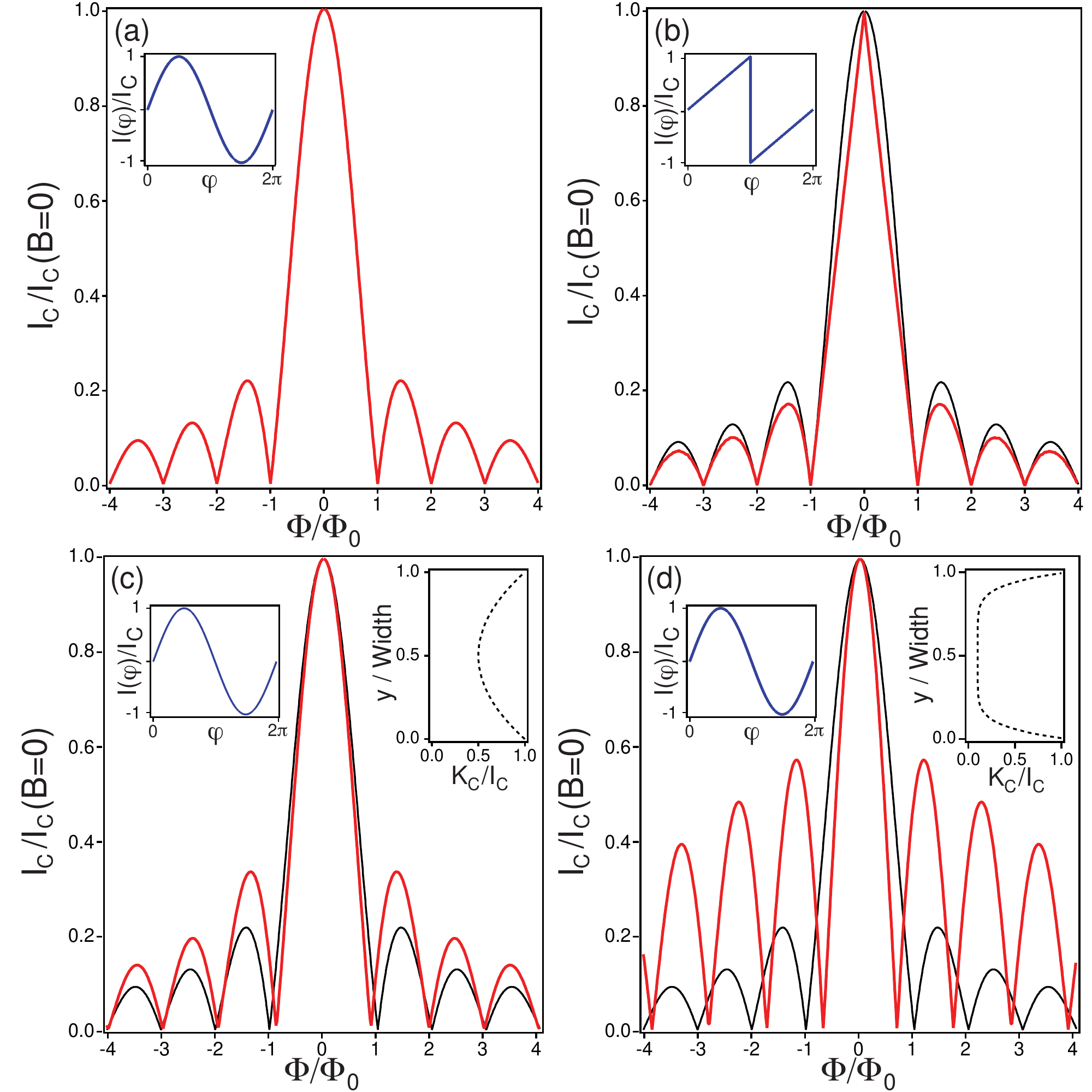}
\caption{Magnetic diffraction patterns (red) for (a) sinusoidal $I(\varphi)$, (b) saw-toothed sinusoidal $I(\varphi)$, (c,d), sinusoidal  $I(\varphi)$ with two different nonuniform critical current line densities $K_c$. In (b-d), the magnetic diffraction pattern for the sinusoidal $I(\varphi)$ (black) from (a) is plotted for comparison. In each figure, the current-phase relation used to produced the magnetic diffraction pattern is shown in the inset.}
\end{figure}

\subsection{Extraction of the Velocity of Dissipative Excitations}
Using the phenomenological model of the main text, the velocity $\nu_{ex}$ of dissipative excitations in the junction can be extracted from a linear fit of the data in Fig. 2(b) assuming a unity proportionality constant between $\icrn$ and $\ec$ (i.e. $\icrn$=$\ec$/e). As $\icrn$ is not completely linear in $1/W$, we focus on $1/W<1\,\mu$m$^{-1}$, yielding $\nu_{ex}$=1.4$\pm$0.2$\times$10$^4$\,m/s. Electrical transport measurements of the Fermi velocity $\nu_F$ for surface electrons in Bi$_2$Se$_3$ range from 10$^5$\,m/s~[S18] to 10$^6$\,m/s~[S10], bracketing the value $\nu_{F}$=4.2$\times$10$^5$\,m/s~[S4] extracted from ARPES, and larger than our inferred value. In fact, a lower group velocity is expected for the bound pairs of electrons and holes that shuttle Cooper pairs across the device~[S19]. Specifically for Majorana fermions -- a subset of these bound pairs -- the velocity $\nu_M$ has been predicted to be less than $\nu_F$ by a factor $(\Delta/\mu)^2$ in Ref.~[S15] and $(\Delta/\mu)$ in Ref.~[S20] (calculated for neutral modes created in graphene JJs), where $\mu$ is the chemical potential of the TI weak link relative to the Dirac point of the surface states. The typical ratio for $\Delta/\mu$ in our samples $\sim$10$^{-3}$~[S4], giving an estimate for the Majorana velocity of .42\,m/s for $(\Delta/\mu)^2$ and 4.2$\times$10$^2$\,m/s for $(\Delta/\mu)$, closer to our measured value though still off by a factor of 20 to 30. The discrepancy between the theoretically predicted value of Ref.~[S20] and the velocity extracted from experiment could occur for two reason. First, the proportionality constants in the relationship between $\icrn$ and $\ec$ and between $\nu_M$ and $\nu_F$ are not known. Second, there is no direct measure of the chemical potential of the surface state in our devices. For samples of density similar to ours, it has been shown that the chemical potential of the surface is less than that of the bulk due to band bending~[S4]. This reduction of the surface chemical potential will cause an increase in the expected Majorana velocity. 

\subsection{Simulation of Josephson Effect in the Presence of Magnetic Flux}
The critical current through the devices as a function of applied field was calculated in a manner closely following that of Tinkham~[S5], but with the extended junction model modified to allow non-standard current-phase relationships and spatial inhomogeneities in critical current density. With the flake surface parallel to the $x-y$ plane, applied field $B$ along $z$, and current along $x$, the phase difference between the leads as a function of the position y along the junction is given by
\begin{equation}
\varphi(y) = \varphi_0 + \frac{2\pi}{\Phi_0}\int_0^y dy'\int_0^{L+2\lambda_L}dx B_z(x,y'),
\end{equation}
where $\varphi_0$ is the phase difference at y=0, $\Phi_0$ is the magnetic field flux quantum, $L$ is the length of the junction, and $\lambda_L$ is the London penetration depth. This follows from integrating the vector potential \textbf{A} around a rectangular contour that includes a path at y'=0 and y'=y and taking advantage of the fact that \textbf{A} = ($\Phi_0/2\pi)\nabla\gamma$ inside the superconducting leads, where $\gamma$ is the non-gauge-invariant phase [defined so that $\varphi=\Delta\gamma-(2\pi/\Phi_0)\int\textbf{A}\cdot d\textbf{s}$]. With normalized current-phase relationship (CPR) $i(\varphi)=I(\varphi)/I_C$ and critical current line density $K_c(y)$, we obtain critical current
\begin{equation}
\ic = max_{\varphi_0}\int_0^W dy K_c(y) i(\varphi(y)).
\end{equation}
Note that this analysis is equally valid for 2$\pi$- and 4$\pi$-periodic phase relations, so long as for the latter case we allow $\varphi(y)$ to range up to 4$\pi$ and define $i(\varphi)$ over the full range. For a typical 2$\pi$-periodic current-phase relationship $I(\varphi)$ = $\ic sin(\varphi)$, and uniform current density and field, the critical current drops to zero when an integer number of flux quanta are threaded through the junction area, corresponding to an integer number of cycles of Josephson current. A simulation of magnetic diffraction pattern (MDP) for $i(\varphi) = sin(\varphi)$ shows the typical Fraunhofer pattern~[S5] [Fig. S3(a)]. Deviations from this typical pattern can occur because of non-sinusoidal CPR or from nonuniform critical current distribution along the junction. For a ballistic JJ, the CPR becomes more of a saw-toothed shape, for which the corresponding MDP is shown in Fig. S3(b). There is a small change in the shape of the MDP, but the minima still occur at integer multiples of $\Phi_0$. Aside from a changing CPR, deviations from the Fraunhofer pattern for short junctions can occur for a non-uniform current distribution. Reductions of the flux at first minimum as large as a factor of 2 can occur in the extreme limit when all of the current is concentrated near the edges of the junction~[S21]. The MDP in this extreme case is shown for two different current distributions in Fig. S3(c,d). None of these deviations produces a MDP similar to Fig. 3 of the main text. 

\begin{figure}[t!]
\center \label{figs4}
\includegraphics[width=3 in]{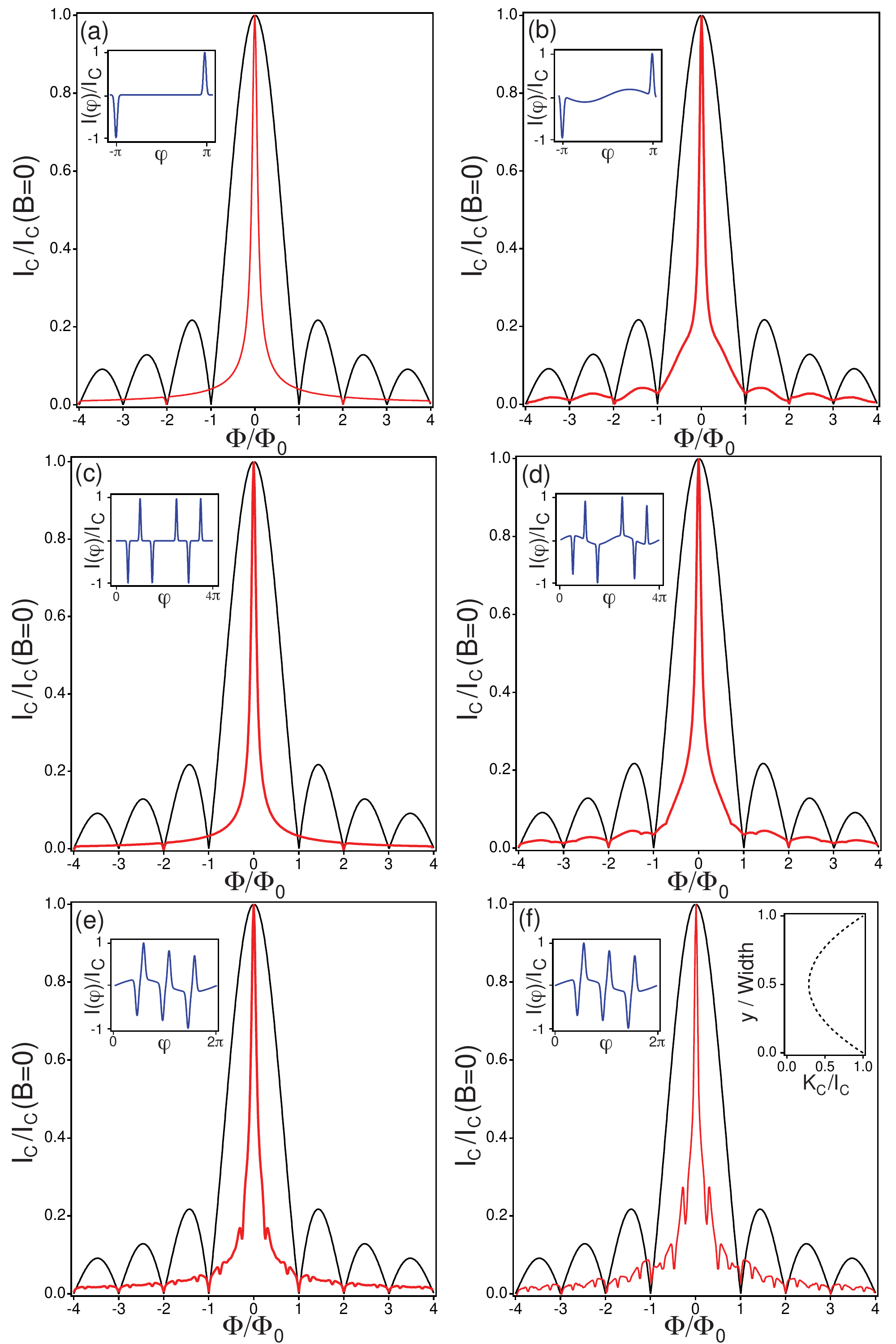}
\caption{Magnetic diffraction patterns (red) for (a) two gaussian peaks in $I(\varphi)$ at $\varphi = \pm \pi$. (b) the same as (a) but with an added conventional sinusoidal $I(\varphi)$ of 1/5 the amplitude of the anomalous peaks. (c), The same as (a), with additional peaks at $\varphi=\pi/3, \pi/2$ and $4\pi/3, 3\pi/2$, resulting from additional zero-energy crossings.  (d) the same as (c) but with a conventional sinusoidal $I(\varphi)$ of 1/5 the amplitude of the anomalous peaks. In each figure, the current-phase relations used to produced the magnetic diffraction pattern -- all 4$\pi$-periodic -- are shown in the inset, and each diffraction pattern is compared to the conventional sinusoidal case (black). (e) (inset) A CPR arising from an ABS spectrum that does not conserve fermionic parity in the junction and contains peaks at $\pi/2, \pi, 3\pi/2$. (main) MDP resulting from the CPR of the inset, produces a sub-$\Phi_0$ structure reminiscent of that seen in experiment. (f) same as (e), except with a nonuniform current distribution (inset) used to more closely reproduce the experimental results.}
\end{figure}

Using this simulation approach, we empirically attempt to determine a CPR to match the MDP we observe in our devices. Not any CPR is possible: the CPR must be antisymmetric $I(\varphi)=-I(-\varphi)$, it must come from the derivative of the energy of the junction and so must average to zero over a period such that the energy of the junction $E(\varphi) \propto \int d\varphi I(\varphi)$ does not continue to grow as a function of $\varphi$~[S22]. In addition, the CPR must contain a term for the contribution of the bulk, which was estimated in our devices to be smaller than the surface contribution, but not zero. In the following, we assume the bulk has a CPR $I(\varphi) \propto sin(\varphi)$. To begin, we assume the simplest CPR that is peaked at values of $\pi$ and 3$\pi$, with alternating signs but equal amplitude to satisfy the conditions on the CPR. This result is shown in Fig. S4(a), where we see that the inclusion of peaks at $\varphi=\pi, 3\pi$ produces a narrowed MDP. It does not, however, reproduce the additional minima seen in Fig. 3. Note that this CPR is 4$\pi$-periodic, as expected for TI JJs with constant Majorana occupancy. The first correction to the simple, peaked CPR above comes from the addition of the bulk term, which we include as a simple sinusoidal dependence on $\varphi$, $sin(\varphi)$, with 1/5 the amplitude (as estimated above) of the peaks in the CPR (S4B). Comparing to S4B to S4A, additional features in the MDP appear in S4B, are thereby introduced at $n\Phi_0$, where $n$ is an integer.  Maintaining only a single strong peak and dip in a 2$\pi$ period of CPR cannot produce a minimum in MDP below $\Phi$=$\Phi_0$. To get closer to the observed MDP, we add more zero-energy crossings, predicted in Ref.~[S17]. We add two additional zero-energy modes or either side of $\varphi=\pi$ and $3\pi$, at $\varphi=\pi/2, 3\pi/2$ and $5\pi/2, 7\pi/2$, and the corresponding MDP is shown in Fig. S4(c,d). The MDP is calculated without [S4(c)] and with [S4(d)] the $sin(\varphi)$ contribution from the bulk.  In Fig. S4(d), a small notch in the MDP appears below the first expected minima, occurring at a value of 3$\Phi_0$/4. A CPR similar to that used in Fig. S3(c,d) had previously been suggested as an anomalous contribution to the CPR from Majorana fermions~[S16], a result of difference in Josephson current depending the even versus odd Majorana occupancies in the JJ. The last CPR considered are of the form from Fig. S2(h) -- assuming the system always relaxes to the ground state -- and are shown in Fig. S4(e,f), where we again have used the multiple zero-energy crossings of Ref.~[S17] but now added a nonuniform current distribution that may occur in our junctions due to supercurrent carried by the sides of our device (Note, the sides of the topological insulator are also expected to have a surface state, yet any supercurrent carried by the sides will be unaffected by the applied magnetic field). The MDP of Fig. S4(f) produces minima at values $\Phi_0/4$, $\Phi_0/2$ and $\Phi_0$, near the experimentally observed minima. It is not possible even in principle to determine a unique CPR from the diffraction pattern but can only supply evidence for a peaked CPR. In our simulations, we were unable to obtain sub-$\Phi_0$ dips for any conventional (i.e. non-peaked) CPR, but were only able to obtain MDPs that look similar to our experimental results using peaked CPRs, with multiple peaks over the range $\varphi$=0 to 2$\pi$. Even in these cases, we do not get a precise match to the data -- we have tried to limit the number of fitting parameters. 

\begin{figure}[t!]
\center \label{figs5}
\includegraphics[width=3 in]{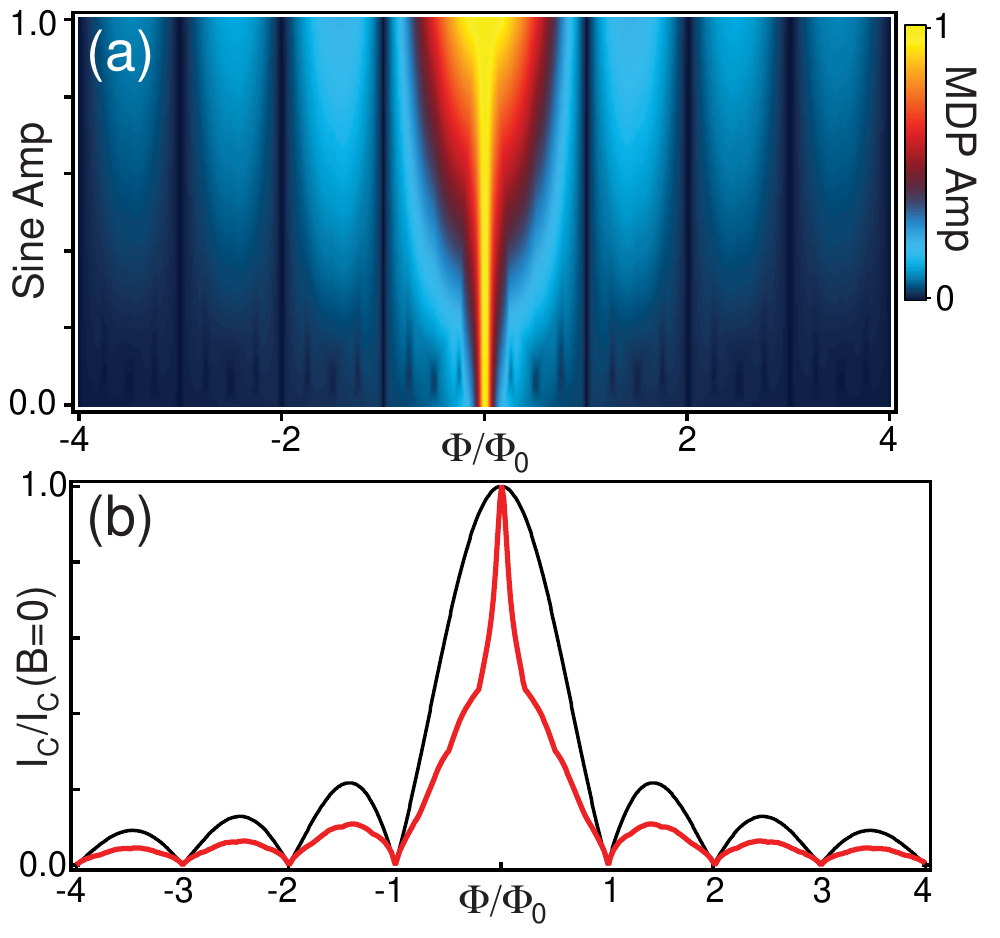}
\caption{(a) A 2D plot of the MDP as a function of $\Phi$ and the amplitude of the sinusoidal bulk contribution (Sine Amp). Sub-$\Phi_0$ minima are observed for values of Sine Amp less than $\sim$0.4. (b) A 1D cut in (a) at a value of Sine Amp=0.5 (red), showing a narrowed central lobe [compared to the Fraunhofer pattern (black)] in the MDP but no minima below $\Phi_0$.}
\end{figure}

MDPs for a varying bulk contribution are shown in Fig. S5. Fig. S5(a) is a 2D plot of the MDP for the CPR of Fig. S4(e) as a function of $\Phi$ and the amplitude of the sinusoidal contribution (Sine Amp).  When Sine Amp=0 (all surface), the resulting MDP is the same as in Fig. S4(e), while for Sine Amp=1 (all bulk) the MDP is same as Fig. S3(a). Fig. S5(a) shows that sub-$\Phi_0$ minima are obtained for values of Sine Amp less than $\sim$0.4. For Sine Amp=0.5 (red), the MDP obtained is shown in Fig. S5(b), where it is seen that a narrowing of the central feature occurs when compared to a Fraunhofer pattern (black), but the minima still occur at integer values of $\Phi_0$. This may account for prior reports finding a relatively conventional MDPs for JJs on TIs. For example, a MDP similar to the one calculated in Fig. S5(b) is seen in Fig. 2(b) of Ref.~[S23]. 

\begin{figure}[t!]
\center \label{figs6}
\includegraphics[width=3 in]{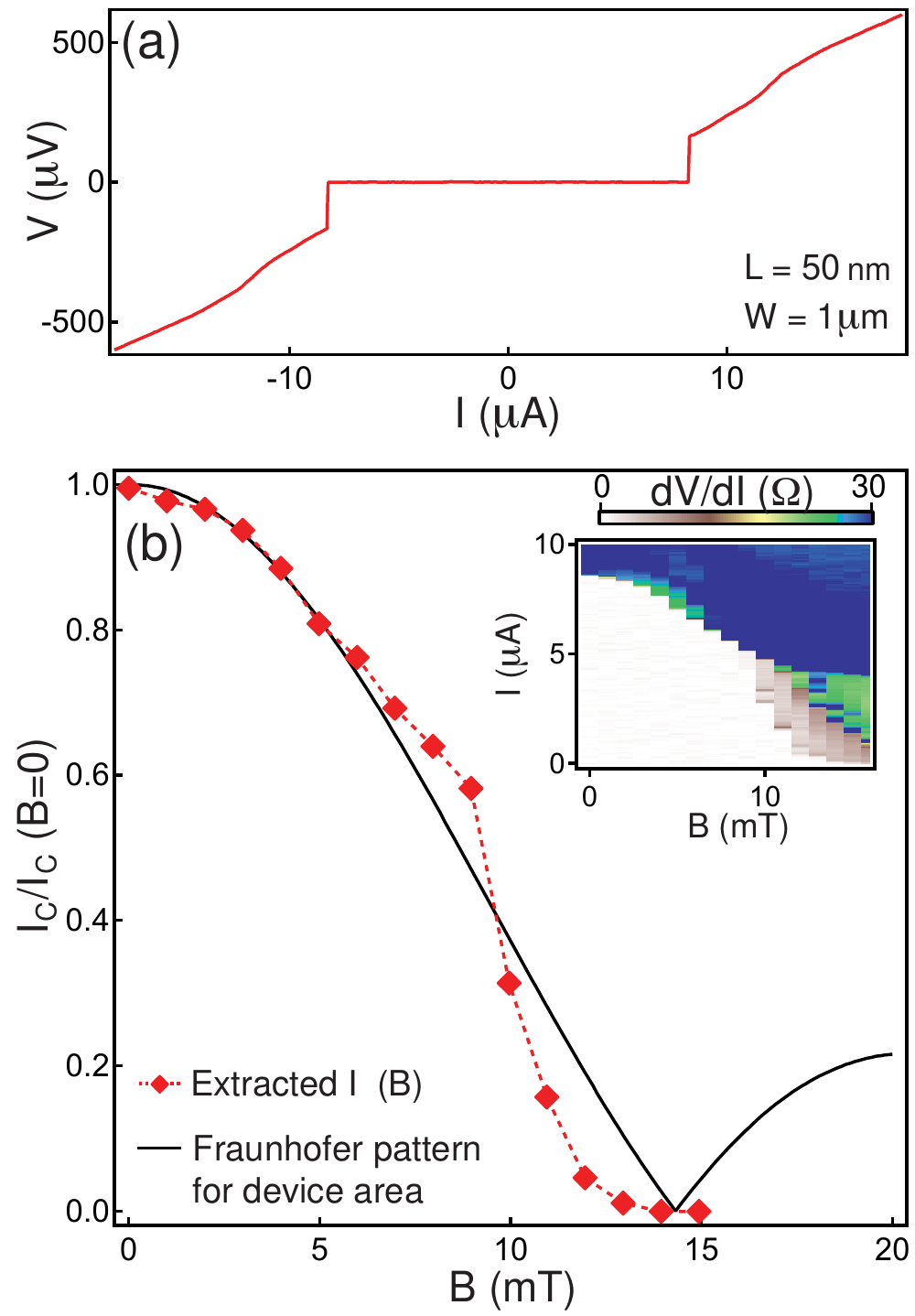}
\caption{(a) $\vdc$\,vs.\,$\idc$ for the ($L,W$)=(50\,nm,1\,$\mm$) control device created from a 75\,nm-thick graphite weak link, producing a $\icrn$ of 244\,$\mu$V, far larger than that observed in TI JJs. (b) Extracted $\ic$ vs. $B$ (red) from $dV/dI$ (inset) giving a value of $\bc$ of 13\,mT, close to the value of 14.3\,mT expected from the device area.}
\end{figure}

\subsection{A Graphite Control Device}
To confirm that the anomalous features observed in TI JJs are a result of the interface between a TI and a conventional superconductor, a JJ was created using 75\,nm-thick graphite as the weak link. The details of the device fabrication are identical to the fabrication of TI JJs, except the Ar ion mill step was not needed to produce low contact resistance. Instead, a 5\,min UV-ozone exposure, typical in the creation of graphene devices~[S24], was used. The results, taken at $T_f$=12\,mK, are shown in Fig. S6 where it is seen that the value of $\icrn$ is much higher at 244\,$\mu$V [Fig. S6a], close to the predicted value of 281\,$\mu$V and 427\,$\mu$V, and $\bc$ is $\sim$13\,mT, closer to the predicted value of 14.3\,mT (black line) for a device of ($L,W$)=(0.05,1)\,$\mm$.  The control device allows for the exclusion of two factors that might have caused the anomalous results in TI JJs. In Fig. S6(a), a plot of $\vdc$ vs. $\idc$ shows a value for the critical current of 8.1\,$\mu$A on a device of resistance 30.2 $\Omega$. The larger value of $\icrn$ obtained for the control device rules out the possibility of the thermal effects caused by poor filtering being a source of the reduced $\icrn$ in the TI JJs. In Fig. S6(b), $\ic$ vs. $B$ (red), extracted from a plot of $dV/dI (B,\idc)$ (inset of S6B) shows a rather conventional dependence on $B$, closely matching the expected pattern (black). The high value of $\bc$ in the control device rules out the possibility of flux focusing~[S25] reducing the values of $\bc$ in TI JJs. $\bc$ for the leads is about 13\,mT for this device, so we cannot measure more lobes in the MDP.

\section*{Supporting References and Notes}
\begin{enumerate}

\item[S1.] A. K. Geim, K. S. Novoselov, \emph{Nature Mater.} \textbf{6}, 183-191 (2007).

\item[S2.] R. M. Potok, Thesis, Harvard Univesity (2006).

\item[S3.] J. G. Analytis, \emph{et al.}, \emph{Nature Phys.} \textbf{6}, 960-964 (2010). 

\item[S4.]  J. G. Analytis, \emph{et al.}, \emph{Phys. Rev. B} \textbf{81}, 205407 (2010).

\item[S5.] M. Tinkham, \emph{Introduction to Superconductivity}. (Dover Publications, Mineola, New York 1996).

\item[S6.] K. K. Likharev, \emph{Rev. Mod. Phys.} \textbf{51}, 101-159 (1979).

\item[S7.]  D.-X. Qu \emph{et al.}, \emph{Science} \textbf{329}, 821 (2010).

\item[S8.] K. K. Likharev, \emph{Dynamics of Josephson Junctions and Circuits}. (Gordon and Breach Science Publishers, Amsterdam 1986).

\item[S9.] G. E. Blonder, M. Tinkham, T. M. Klapwijk, \emph{Phys. Rev. B} \textbf{25}, 4515-4532 (1982). 

\item[S10.] B. Sac\'ep\'e, \emph{et al.}, \emph{Nature Comm.} \textbf{2} 575, (2011).

\item[S11.] A. Furusaki, \emph{Superlat. and Microstruct.} \textbf{25}, 809-819 (1999). 

\item[S12.] H.-J. Kwon, K. Sengupta, K., V. M. Yakovenko, \emph{Euro. Phys. J. B} \textbf{37}, 349-361 (2003).

\item[S13.] L. Fu, C. L. Kane, \emph{Phys. Rev. B} \textbf{79}, 161408 (2009).

\item[S14.] C. L. Kane, M. P. A. Fisher, \emph{Phys. Rev. Lett.} \textbf{68}, 1220-1223 (1992).

\item[S15.] L. Fu, C. L. Kane, \emph{Phys. Rev. Lett.} \textbf{100}, 096407 (2008).

\item[S16.] P. A. Ioselevich, M. V. Feigel'man, \emph{Phys. Rev. Lett.} \textbf{106}, 077003 (2011).

\item[S17.] K. T. Law, P. A. Lee, \emph{Phys. Rev. B} \textbf{84}, 081304 (2011).

\item[S18.] M. Veldhorst, \emph{et al.}, \emph{arXiv:112.3527} (2011).

\item[S19.] A. V. Shytov, P. A. Lee, L. S. Levitov, \emph{Phys. Usp.} \textbf{41}, 207-210 (1998).

\item[S20.] M. Titov, A. Ossipov, C. W. J. Beenakker, \emph{Phys. Rev. B} \textbf{75}, 045417 (2007).

\item[S21.] A. Barone, G. Patern\`o, \emph{Physics and Applications of the Josephson Effect}. (Wiley-Interscience Publications, Canada 1982).

\item[S22.] A. A. Golubov, M. Y. Kupriyanov, E. Il'ichev, \emph{Rev. Mod. Phys.} \textbf{76}, 411-469 (2004).

\item[S23.] F. Qu, \emph{et al.}, \emph{arXiv:112.1683} (2011). 

\item[S24.] J. R. Williams, L. Dicarlo, C. M. Marcus, \emph{Science} \textbf{317}, 638-641 (2007).

\item[S25.] M. B. Ketchen,  W. J. Gallagher,  A. W. Kleinsasser, S. Murphy, J. R. Clem, in \emph{Proceedings of SQUID Õ85} (Berlin 1985) H. D. Hahlbahm, H. Lubbig, Eds., pp. 865.

\end{enumerate}


\begin{thebibliography}{References}
\bibitem{SuppInfo} Information on materials and methods will be available online.
\end{thebibliography}
\end{document}